**FRONT MATTER**

## Title
- Impurity contribution to ultraviolet absorption of saturated fatty acids
- Impurities in photoabsorption of fatty acids

## Authors

Shota Saito,[1] Naoki Numadate,[1] Hidemasa Teraoka,[1] Shinichi Enami,[2] Hirokazu Kobayashi,[1] Tetsuya Hama[1*]

## Affiliations

1 Komaba Institute for Science and Department of Basic Science, The University of Tokyo, Meguro, Tokyo 153-8505, Japan.
2 Department of Chemistry, Graduate School of Science and Technology, University of Tsukuba, Tsukuba 305-8571, Japan.

*Corresponding author. Email: hamatetsuya@g.ecc.u-tokyo.ac.jp

## Abstract

Saturated fatty acids are abundant organic compounds in oceans and sea sprays. Their photochemical reactions induced by solar radiation have recently been discovered as an abiotic source of volatile organic compounds, which serve as precursors of secondary organic aerosols. However, photoabsorption of wavelengths longer than 250 nm in liquid saturated fatty acids remains unexplained, despite being first reported in 1931. Here we demonstrate that the previously reported absorption of wavelengths longer than 250 nm by liquid nonanoic acid [$CH_3(CH_2)_7COOH$] originates from traces of impurities (0.1% at most) intrinsically contained in nonanoic acid reagents. Absorption cross sections of nonanoic acid newly obtained here indicate that the upper limit of its photolysis rate is three-to-five orders of magnitude smaller than those for atmospherically relevant carbonyl compounds.

## Teaser

Impurities in saturated fatty acids, rather than the acids, absorb ultraviolet wavelengths present in sunlight at Earth's surface.

## MAIN TEXT

**The manuscript should be a maximum of 15,000 words.**

## Introduction

Sunlight is Earth's largest source of energy. Absorption of sunlight by molecules initiates photochemical processes that determine many aspects of atmospheric and climate chemistry. The rates of photolysis of molecules in the atmosphere are thus fundamental quantities for further understanding local and global climate change and also public health effects (*1*).

Photolysis rates depend not only on the intensity of solar irradiation, but also on the photochemical and photophysical properties of the absorbing molecules, such as the absorption cross section. The study of gas-phase homogeneous reaction systems in the atmosphere has evolved dramatically since the Chapman theory in the 1930s first explained the stratospheric ozone layer (1, 2). Databases now contain cross sections for the absorption of ultraviolet (UV) wavelengths by many gaseous molecules and radicals relevant to tropospheric and stratospheric chemistry (3–5).

A challenge for current research in atmospheric chemistry is to expand our understanding of multiphase chemistry; i.e., chemical reactions involving transport and transformations among gases, liquids, and solids (2, 6–10). For example, photochemical reactions of liquid saturated fatty acids (i.e., acyclic aliphatic carboxylic acids (11)) have attracted much research attention because they are abundant in natural environments such as sea-surface microlayers and atmospheric organic aerosols (12–16). Laboratory studies have shown that direct UV photolysis of liquid-phase nonanoic acid [$CH_3(CH_2)_7COOH$], a representative fatty acid present in sea-surface microlayers and sea sprays, leads to the formation of oxidized volatile organic compounds (e.g., aldehydes and ketones) in the gas phase (6, 17, 18). These volatile organic compounds are proposed to serve as precursors for the formation of secondary organic aerosols in the troposphere (19–21), and they eventually influence climate through the formation of cloud condensation nuclei (22).

Quantitative estimation of the photolysis rate of a saturated fatty acid in the troposphere requires accurate absorption cross sections, especially for wavelengths longer than 295 nm (23–25). However, the absorption cross sections of liquid molecules, including simple fatty acids such as nonanoic acid, are subject to debate. UV absorption spectra of liquid aliphatic carboxylic acids often exhibit a weak shoulder absorption band centered at 270 nm and extending to 330 nm, in addition to the main absorption band at around 210 nm, due to a singlet–singlet ($S_0$–$S_1$) $n{\rightarrow}\pi^*$ transition (6, 26–28). Figure 1A(a) shows the UV absorption spectrum of nonanoic acid reported by Rossignol et al. (see also Fig. S1) (6). Since the first report of UV absorption by fatty acid in 1931 (26), various origins have been proposed for the shoulder absorption; e.g., carboxylate anions (26), a singlet–triplet ($S_0$–$T_1$) $n{\rightarrow}\pi^*$ spin-forbidden transition of neutral molecules (6, 27), and the formation of cyclic dimers.(17) However, no consensus has yet been reached, inhibiting our understanding of the photochemistry of liquid carboxylic acids and evaluation of their impact on tropospheric chemistry.

An intrinsic problem facing the acquisition of absorption spectra for liquid samples is the contribution of impurities (29, 30). Commercially available chemical reagents are often assumed to be sufficiently pure for most experimental studies. However, much discussion has considered the possibility that undetectable trace amounts of impurities (micromolar or nanomolar concentrations) greatly affect experimentally observed phenomena (31–38). Impurities can particularly influence measurement of the UV absorption spectra of liquid samples because a large volume of sample is usually required: a typical quartz cuvette cell with an optical path length of 10 mm holds 3.5 mL. Arudi et al. found that impurities in unsaturated fatty acids (oleic and linoleic acids) contribute markedly to the UV absorption observed around 270 nm, even for samples of the highest purity (>99.0%) (39). This suggests that the contribution of impurities to the UV absorption of saturated fatty acids should be carefully evaluated.

**Results**

This study experimentally demonstrated that the weak shoulder absorption from 250 to 330 nm observed for a saturated fatty acid (nonanoic acid) originates from traces of impurities (0.1% at most). Figure 1A(b) shows a UV absorption spectrum of commercial nonanoic acid (purity > 98.0%) before purification. The spectrum is essentially identical to that reported by Rossignol et al. (Fig. 1A(a), see also Fig. S1) (6). To evaluate the contribution of impurities to the weak shoulder absorption, we developed a device for recrystallization at low temperature and under anaerobic conditions (Fig. S2) and purified the nonanoic acid by recrystallization 15 times. The UV absorption spectrum of the purified nonanoic acid (Fig. 1A(c)) shows the almost complete disappearance of the weak shoulder absorption. The removal of the impurities required 10–15 recrystallizations, and further recrystallization did not significantly improve the purity of the nonanoic acid (Fig. S3).

To obtain quantitatively reliable absorption cross sections of the purified nonanoic acid over a wide wavelength range (190–310 nm) requires avoidance of saturation of the absorption. Therefore, we collected UV absorption spectra with different optical path lengths of 90, 40, 10, 4, and 1 mm using quartz cuvette cells and of 0.564, 0.105, and 0.0185 mm using demountable liquid cells (Figs S4–S8 and Table S1). These eight spectra provided accurate absorption cross sections of liquid nonanoic acid at 190–310 nm. The cross sections differed by six orders of magnitude, ranging from $10^{-19}$ to $10^{-24}$ cm$^2$, as shown Fig. 1B(c) (see the supplementary materials for details). Table S2 summarizes the values of absorption cross sections with estimated errors. The peak absorption cross section at 205 nm did not differ significantly before or after recrystallization ($2.6 \times 10^{-19}$ vs. $2.4 \times 10^{-19}$ cm$^2$), indicating that photoabsorption at 205 nm was governed by nonanoic acid molecules with insignificant contribution from impurities. However, the absorption cross sections at wavelengths larger than 250 nm decreased drastically following purification by recrystallization (Fig. 1B(b) and (c)). For example, that at 295 nm was $1.3 \times 10^{-23}$ cm$^2$ after purification, 24 times smaller than that before ($3.1 \times 10^{-22}$ cm$^2$; Table S2).

Absorption at wavelengths > 250 nm was clearly related to impurities in the nonanoic acid (Fig. 1B). The absorption cross section at 295 nm ($1.3 \times 10^{-23}$ cm$^2$) was three orders of magnitude smaller than those typical for atmospherically relevant organic molecules such as formaldehyde ($H_2CO$, $4.4 \times 10^{-20}$ cm$^2$), acetaldehyde ($CH_3CHO$, $4.3 \times 10^{-20}$ cm$^2$), and acetone ($CH_3(CO)CH_3$, $3.4 \times 10^{-20}$ cm$^2$) at 295 nm (1 nm average) in the gas phase (Table S3–S5) (3–5). Therefore, compared with these carbonyl compounds, nonanoic acid shows negligibly small photoabsorption of wavelengths present in the troposphere. The absorption cross sections of nonanoic acid at wavelengths > 250 nm obtained here represent upper limits, as trace amounts of impurities can remain even after 15 recrystallizations, as shown by the UV absorption spectrum with an optical path length of 90 mm (Fig. S5). We also confirmed that octanoic acid ($CH_3(CH_2)_6COOH$, purity ≥ 99%) contained impurities with a much stronger contribution to its absorption of wavelengths > 250 nm than seen for nonanoic acid (purity > 98.0%), despite its higher purity (Fig. S9). This suggests that impurities generally affect UV absorption by saturated fatty acids.

To identify the chemical characteristics of the impurities, we also analyzed the nonanoic acid before recrystallization using high-performance liquid chromatography (HPLC) with a photodiode array (PDA) detector. This detected at least seven impurities with significant UV absorption peaks at 250–300 nm (Fig. S10), implying that the impurities had carbonyl or conjugated structures. We also measured the nuclear magnetic resonance (NMR) spectra of nonanoic acid. NMR is usually unsuitable for the detection of small amounts of impurities, but an 18.8 T cryo-probe NMR could detect impurities at ~0.1%. Figure 2 shows $^{13}$C NMR spectra of nonanoic acid (purity > 98.0%) at 50% concentration in deuterated chloroform ($CDCl_3$, purity 99.8%) solutions before and after recrystallization. Before recrystallization, the sample yielded many peaks originating from impurities in addition to those for nonanoic acid (180.6 ppm) and $CDCl_3$ (77.2, 77.0, and 76.9 ppm) (Fig. 2A) (40). Measurement of $CDCl_3$ alone confirmed it had no clear impurity peaks (Fig. S11). The intensities of impurity peaks were approximately 0.1%–3% of those of nonanoic acid. After recrystallization, these impurity peaks were almost completely removed, leaving only the peaks for nonanoic acid, $CDCl_3$, and the acetonitrile ($CH_3CN$; 115.9 ppm) solvent used for recrystallization (Figs 2B and S12). Two-dimensional NMR (2D-NMR) clarified that the most intensive impurity peak in Fig. 2A (183.7 ppm; ~3% of the nonanoic acid peaks) was assigned to a carboxylic acid with a methyl (–$CH_3$) group at the α-position (Fig. S13). Considering its persistence during recrystallization, a candidate for this most abundant impurity is 2-methyloctanoic acid [2-MOA, $CH_3(CH_2)_4CH(CH_3)COOH$]. However, it is unlikely that such a saturated aliphatic compound absorbs light with wavelengths longer than 260 nm, given its structural similarity to nonanoic acid. As HPLC–PDA also indicated no absorption peak >260 nm for this compound (Fig. S10), the UV absorption was derived from minor impurities, which is why this critical issue has been overlooked in previous studies.

Although it was difficult to identify precise molecular structures for other minor impurities (showing peaks about 0.1% of that of nonanoic acid at 180.6 ppm), their peak positions indicated their functional groups (Fig. 2A). The peaks at 210.4 and 208.8 ppm were attributed to the carbonyl carbons in ketones and/or aldehydes (41). Other peaks indicated carboxylic groups (e.g., 172.6 and 179.4 ppm), alkenes and/or aromatics (e.g., 156.0 and 121.3 ppm), and aliphatic C–O groups such as alcohols and ethers (at 90 to 60 ppm) (41). Analysis by 2D-NMR clarified that the peaks at 208.8 and 179.4 ppm originated from a keto acid (Fig. S14). Among these impurities, ketones including keto acids and/or aldehydes can absorb photons at 240–340 nm via $n \rightarrow \pi^*$ transitions with absorption cross sections of the order of $10^{-20}$ $cm^2$ in the gas phase (1, 3, 4) and thus contribute to the UV absorption of nonanoic acid before purification (Fig. 1A).

## Discussion

Our results indicate that, before recrystallization, nonanoic acid's absorption of UV wavelengths > 250 nm is dominated by impurities at concentrations of ~0.1% at most, rather than by carboxylate anions, cyclic neutral dimers, or singlet–triplet ($S_0$–$T_1$) $n \rightarrow \pi^*$ spin-forbidden transitions of neutral molecules, as proposed in previous studies.(6, 17, 26, 27) Quantum calculations supported this conclusion, as none of the structures of carboxylic acids (monomers, hydrated structures, dimers, deprotonated anions, or protonated cations) have absorption peaks above 250 nm (Fig. S15 and Table S6). The new absorption cross sections of nonanoic acid obtained here have several implications for atmospheric chemistry, especially in the troposphere. The rate of atmospheric photodissociation of nonanoic acid, $J$ ($s^{-1}$), is given by equation (1):

$$J = \int \sigma(\lambda,T)\Phi(\lambda,T)F(\lambda,z,\chi)d\lambda \tag{1}$$

where $\sigma(\lambda, T)$ is the absorption cross section (cm$^2$) of nonanoic acid at wavelength $\lambda$ and temperature $T$, $\Phi(\lambda, T)$ is the quantum yield for photodissociation at $T$, and $F(\lambda, z, \chi)$ is the actinic flux (photon cm$^{-2}$ nm$^{-1}$ s$^{-1}$), which is a function of altitude $z$ and solar zenith angle $\chi$ (42–44). The integration is carried out over the wavelength region of interest. Assuming a quantum yield of unity for photolysis (i.e., $\Phi(\lambda, T) = 1$), the upper limit of the photolysis rate ($J$-value) of nonanoic acid at room temperature is estimated to be $1.0 \times 10^{-9}$ s$^{-1}$ in the troposphere, given reported values of the actinic flux at wavelengths of 292–310 nm at 0.1 km altitude and a solar zenith angle of 17° (Fig. 3 and Table S7).(23) This upper limit is respectively 660, 5000, and 98,000 times smaller than those calculated for the atmospherically important carbonyl compounds acetone (CH$_3$C(O)CH$_3$, $6.6 \times 10^{-7}$ s$^{-1}$ at 292–327 nm), acetaldehyde (CH$_3$CHO, $5.0 \times 10^{-6}$ s$^{-1}$ at 292–332 nm), and formaldehyde (H$_2$CO, $9.8 \times 10^{-5}$ s$^{-1}$ at 292–361 nm) in the gas phase (Tables S3−S5). There are no accurate measurements of the absorption cross section at wavelengths > 295 nm for carboxylic acid molecules in the gas phase (including nonanoic acid) because of their weak absorption (3, 4, 45, 46). The present upper limit of the absorption cross section of liquid nonanoic acid would facilitate proper evaluation of the photoabsorption of carboxylic acids in atmospheric chemistry models (21). Although Rossignol et al. proposed that a monolayer of nonanoic acid at the air–water interface weakly absorbs UV wavelengths present in sunlight at Earth's surface (6), their argument was essentially based on the weak shoulder absorption around 270 nm shown by neat or concentrated solutions of nonanoic acid without purification. The present revelation that this absorption originates from impurities necessitates reevaluation of previous experimental studies about the photochemistry of fatty acids, both in the neat liquid phase and in monolayers at air–water interfaces, with careful consideration of the effects of impurities. For example, Rossignol et al. proposed the direct dissociation of nonanoic acid to form acyl (RCO) and hydroxyl (OH) radicals as a possible photochemical reaction mechanism (6). However, the photodissociation of ketone impurities (RCOR') at around 300 nm can also generate RCO radicals and OH radicals can be subsequently produced from the RCO + O$_2$ reaction in the presence of oxygen molecules (O$_2$) (47).

Recent advances in analytical methods have increased the sensitivity of detecting trace amounts of chemicals including impurities and external contaminants.(32, 48, 49) Although complete elimination of impurities and contaminants may be infeasible, more precise evaluation and control are necessary to ensure the accuracy and reliability of scientific research. This is particularly important in research of photochemical systems because impurities and contaminants can act as photosensitizers when they are selectively excited by photons.

## Materials and Methods
### Chemicals

Nonanoic acid was purchased from Tokyo Chemical Industry (TCI, purity > 98.0%) and Thermo Scientific Chemicals (purity 97%). The latter product was identical to the Alfa Aesar product used by Rossignol et al (6). As both showed similar UV absorption spectra, we focused on the TCI reagent for purification (Fig. S1). Acetonitrile (purity ≥ 99.5%, Nacalai Tesque) solvent was used for recrystallization. Octanoic acid (purity ≥ 99%) was from Sigma-Aldrich. Deuterated chloroform (CDCl$_3$, purity 99.8%) was from Fujifilm Wako Pure Chemical Corporation. Purified liquid H$_2$O (resistivity ≥ 18.2 MΩ cm at 298 K) was obtained from a Millipore Milli-Q water purification system.

## Purification method

A liquid mixture of 500 mL nonanoic acid and 1000 mL acetonitrile was placed in a 2 L gas washing bottle (NBO-2L-SCI, Hario; Fig. S2). A 1 L gas washing bottle (014660-1000, Sibata Scientific Technology Ltd.) was also used depending on the sample volume desired for UV absorption measurements. Recrystallization involved immersing the bottle in antifreeze solution (SB-EG, As One Corporation) at −27 ± 1 °C in a thermostatic bath (LTB-250α, As One Corporation). Nitrogen gas (purity 99.998%, Kotobuki Sangyo Co., Ltd.) flowed into the bottle during recrystallization to agitate the liquid mixture and maintain an anaerobic atmosphere. During recrystallization, the inside of the bottle was kept at positive pressure with a constant flow of $N_2$ gas to exclude air. After immersing the bottle into the antifreeze solution for a few minutes, crystals of nonanoic acid appeared as foam-like solids. The mother liquor was removed one hour after the crystals first appeared, and the crystals were melted by heating. To repeat the recrystallization, acetonitrile was re-added to the bottle and the same procedure was performed. The removal of impurities required 10–15 recrystallizations, and further recrystallization did not significantly improve the purity (Fig. S3).

After purification, residual acetonitrile in the nonanoic acid sample was removed by holding the sample at 100 Pa and 40 °C. The final yield of purified nonanoic acid was approximately 80 ml (16%), which was sufficiently large for UV absorption measurement using a 100 mm quartz cuvette with a volume of 35.0 mL (Fig. S4).

## UV absorption spectroscopy

A double-monochromator UV–visible spectrophotometer (UV-2550, Shimadzu) measured the UV absorption spectra of liquid samples at wavelengths of 190–310 nm. A deuterium lamp (L6380, Shimadzu) was used as a broad-band UV light source. A slow scanning speed (160 nm/min) was employed with a resolution of 1 nm. The slit width was set to 2 nm.

The Beer–Lambert law gives a sample's absorbance ($A$) at wavelength $\lambda$ (nm) as follows with respect to the absorption cross section $\sigma$ (cm$^2$) (*50*):

$$A = -\log_{10} \frac{I^s}{I^b} = \frac{\sigma n}{\ln 10} \tag{2}$$

where $I$ is the intensity of the transmitted UV light at wavelength $\lambda$ (nm); superscripts $s$ and $b$ indicate background and sample measurements, respectively; and $n$ (cm$^{-2}$) represents the column density of molecules in the sample. Considering the sample density ($D = 0.9052$ g cm$^{-3}$ at 20 °C for nonanoic acid (*51*)), the mean molecular mass ($M = 158.24$ g mol$^{-1}$ for nonanoic acid (*52*)), and Avogadro's constant ($N_A = 6.0221 \times 10^{23}$ mol$^{-1}$), $n$ can be expressed as follows:

$$n = \frac{N_A D L}{M} \tag{3}$$

where $L$ represents the optical path length (mm). Thus, equation (2) can be rewritten as follows:

$$A = -\log_{10} \frac{I^s}{I^b} = \frac{\sigma N_A D L}{M \ln 10} \tag{4}$$

Equation (4) means that $\sigma$ can be derived from $A$ using equation (5):

$$\sigma = \frac{A M \ln 10}{N_A D L} \tag{5}$$

This study used a double-beam arrangement to measure the $A$ (and thus $\sigma$) of nonanoic acid. UV radiation from the double monochromator was split into two beams, one for background and the other for sample measurements. The first beam passed through a cuvette with a short optical path length ($L^b$) to measure $I^b$, and the latter passed through a cuvette with a long optical path length ($L^s$) to measure $I^s$ (Fig. S4). Both cuvettes contained identical nonanoic acid samples. The difference between $L^b$ and $L^s$ corresponded to the optical path length $L$ in equation (4): i.e., $L = L^s - L^b$. This configuration minimized any undesirable optical interface effects (e.g., reflection at the cuvette–sample interfaces) contributing to $I^b$ and $I^s$, because the refractive indices of the liquids were identical in both the background and sample measurements (1.4322 for nonanoic acid(52)). Therefore, the intensity ratio $\dfrac{I^s}{I^b}$ predominantly reflected absorption by a sample with optical path length $L = L^s - L^b$, which led to accurate measurement of $A$, and thus $\sigma$.

To guarantee a linear relationship between $A$ and $n$ for quantitative analysis, saturation of the absorption by nonanoic acid must be avoided, and UV absorption spectra with $A \leq 1$ are desirable to obtain $\sigma$ from $A$. For this purpose, we used rectangular quartz cuvettes with different optical path lengths of 1, 2, 5, 10, 20, 50, and 100 mm (1/Q/1–1/Q/100, Starna Scientific Ltd.) to change $L^b$ and $L^s$, and UV absorption spectra with $L = 1, 4, 10, 40,$ and 90 mm were collected (Fig. S5). Table S1 lists the combinations of cuvettes used to obtain these path lengths.

Figure S5D shows that absorption saturated at wavelengths less than 240 nm when $L = 1$ mm. Hence, UV absorption spectra with $L < 1$ mm were measured using a liquid film method with demountable liquid cells (162-1200, PIKE Technologies; Figs S4 and S6) (53). A liquid film was prepared by compressing a liquid sample with two quartz windows in a demountable liquid cell (Fig. S6). A Teflon spacer with thicknesses of 0.015, 0.025, 0.1, 0.5, or 1 mm was also inserted between the quartz windows to change $L^b$ or $L^s$ (Table S1), and UV absorption spectra with $L$ $(= L^s - L^b)$ of 0.010, 0.085, and 0.5 mm were collected (Fig. S5).

The UV absorption spectra in Fig. S5 cover a wide range of wavelengths between 190 and 310 nm with $A \leq 1$, which allowed calculation of $\sigma$ by equation (5). However, the liquid film measurements had less accurate values of $L$ than those using quartz cuvettes because the thickness of the Teflon spacer could vary when compressed by the quartz windows. Therefore, the actual optical path lengths ($L'$) were calculated for the liquid film measurements by comparing the absorbance ($A$) values measured at $L = 0.5$ mm using liquid film with those measured at $L = 1$ mm using quartz cuvettes in the spectral range with $0.1 \leq A \leq 1$ (Fig. S7). As a result, $L'$ was calculated as $0.564 \pm 0.014$ mm for UV absorption spectra with $L = 0.5$ mm from equation (4) (Table S1). The $L'$ values measured at $L = 0.085$ mm were then further obtained as $L' = 0.105 \pm 0.004$ mm by comparing the $A$ values with by those measured at $L' = 0.564 \pm 0.014$ mm in the spectral range with $0.1 \leq A \leq 1$. The $L'$ values measured at $L = 0.010$ mm were also similarly obtained as $L' = 0.0185 \pm 0.002$ mm (Fig. S7 and Table S1). Using $L'$ in each UV absorption spectrum, $\sigma$ was obtained from $A$ by equation (5) (Fig. S8 and Table S2). The errors in Table S2 represent the standard deviation (1 sigma) of independent statistical experiments (see also Table S1).

We also confirmed that octanoic acid ($CH_3(CH_2)_6COOH$, Sigma-Aldrich, purity $\geq 99\%$) also contained impurities contributing to its UV absorption at wavelengths longer than 250 nm (Fig. S9). The impurity-induced absorption was much stronger than that seen for nonanoic acid, and 15 repeated recrystallizations did not remove the impurities (Fig. S9). The present study thus focused on nonanoic acid.

High-performance liquid chromatography (HPLC)

Nonanoic acid (TCI, purity 98%) was analyzed using high-performance liquid chromatography with a photodiode array detector (Shimadzu, SPD-M20A; Fig. S10). The column used was a Shim-pack Velox SP-C18 (Shimadzu, 100 mm × ⌀3.0 mm, 2.7 μm), and the eluent was $CH_3CN$ (Nacalai Tesque, HPLC grade) / $H_2O$ (Nacalai Tesque, HPLC grade) = 1/2 (vol/vol) with 2 mM $H_2SO_4$. A 1 μL nonanoic acid sample was directly injected to avoid ghost peaks.

Nuclear magnetic resonance (NMR)

All NMR data were acquired at the NMR Platform in the Graduate School of Pharmaceutical Sciences at the University of Tokyo. Spectra were recorded on a Bruker Avance Neo 800 MHz spectrometer, with a TXO cryoprobe (Figs S11–S14). One-dimensional $^1H$ and $^{13}C$ NMR spectra and two-dimensional NMR (2D-NMR) spectra ($^1H$–$^{13}C$ heteronuclear multiple bond correlation, HMBC) were recorded at room temperature using 5 mm NMR tubes (S-5-800-7, Norell, Inc.). All spectra were processed using TopSpin version 4.2.0 (Bruker).

Theoretical calculations

Gaussian 16 A.03 software was used to calculate the excitation energies using density functional theory (DFT) (54), time-dependent DFT (TDDFT) (55), and symmetry-adapted cluster–configuration interaction (SAC-CI) analysis (56, 57). The B3LYP functional (58, 59) was used for the DFT calculations owing to its high accuracy for organic molecules; the CAM-B3LYP functional (60) was used for TDDFT to incorporate long-range correction to improve the accuracy of the excitation energies. The SAC-CI analysis considered all electrons for electron correlation calculations and chose the "LevelThree" option that sets the lowest thresholds for selecting double excitation operators to maximize accuracy. The chemical structures were optimized at the B3LYP/6-311+G(2df,2p) level prior to excitation calculations.

We initially compared nonanoic acid, propionic acid, and acetic acid using the TD-CAM-B3LYP/6-311+G(2df,2p) level of theory (entries S1–S3 in Table S6). All three acids had almost the same energies calculated for excitations from the $S_0$ to the $S_1$ and $T_1$ states; the energies corresponded to wavelengths of 208–210 and 231–232 nm, respectively.

Therefore, we focused on acetic acid so that we could survey many possible structures and confirm the accuracy using SAC-CI, a highly accurate but time-consuming calculation.

Analysis using DFT (B3LYP/6-311+G(2df,2p), unrestricted open shell) was also employed to calculate the $S_0$ and $T_1$ states for acetic acid. In the conformation optimized for the $S_0$ state, the energy difference between the two systems corresponded to 233 nm for $S_0 \rightarrow T_1$ excitation (entry S4), which was virtually the same as that calculated by TDDFT (232 nm, entry S3). Previous DFT calculation (6) found the vertical transition to correspond to 270 nm. That research assumed the excitation energy to be based on the enthalpy of each state at the optimized structure for the $S_0$ state, which induced a significant error in the calculation for the $T_1$ state in the same geometry, because the harmonic oscillator model required for the calculation was invalid (see also Fig. S15). Instead, we used the electronic energy, which is a major cause of the difference between the results.

To confirm the accuracy of TDDFT, SAC-CI calculations were used to predict the excitation energies for $S_0 \rightarrow T_1$. TDDFT yielded 232 nm, and SAC-CI gave 223 nm (entries S3 and S5). The similarity of the results indicates the good accuracy of the TDDFT calculations, although the actual wavelengths for $S_0 \rightarrow T_1$ are possibly slightly shorter than those predicted by TDDFT.

It is established that carboxylic acids produce dimers. Dimerization shifted the excitation wavelength to 202 nm for $S_0 \rightarrow S_1$ and 220 nm for $S_0 \rightarrow T_1$ (entry S6). As a different methodology to include the effects of multi molecules, we incorporated a solvent effect by the integral equation formalism variant of the polarizable continuum model (IEFPCM), and it provided a similar tendency (entry S7). Hence, the presence of multiple acetic acid molecules increased the excitation energy.

We also considered the possibility that cationic and anionic species ($CH_3COOH_2^+$ and $CH_3COO^-$) contributed to the UV absorption. These ionic species are unstable alone; therefore, dimerization with $CH_3COOH$ was assumed to stabilize them. Their excitation wavelengths for $S_0 \rightarrow T_1$ (entries S8 and S9) were 215 and 230 nm.

In the presence of water, acetic acid may form adducts with water molecules. The calculations for a $CH_3COOH \cdots 3H_2O$ adduct and $CH_3COOH$ with the water solvent effect both blue shifted excitation (entries S10 and S11), compared with that for $CH_3COOH$ under vacuum.

Overall, no absorption peaks > 250 nm were found.

**Acknowledgments**

We thank A. Hofzumahas and C. Zerefos for providing the data for the actinic flux in ref (*23*). This work was performed in part at the NMR Platform in the Graduate School of Pharmaceutical Sciences at the University of Tokyo.

**Funding:**

Japan Society for the Promotion of Science KAKENHI grant JP23H03987 (TH)
Japan Society for the Promotion of Science KAKENHI grant JP21H01143 (TH)
Japan Society for the Promotion of Science KAKENHI grant JP22K18019 (NN)

**Author contributions:**

Conceptualization: SS, NN, TH
Methodology: SS, NN, HK, TH
Investigation: SS, NN, HT, HK, TH
Visualization: SS, NN, HK, TH
Supervision: TH
Writing—original draft: SS, NN, HT, SE, HK, TH
Writing—review & editing: SS, NN, HT, SE, HK, TH

**Competing interests:** Authors declare that they have no competing interests.

**Data and materials availability:** All data are available in the main text or the supplementary materials.

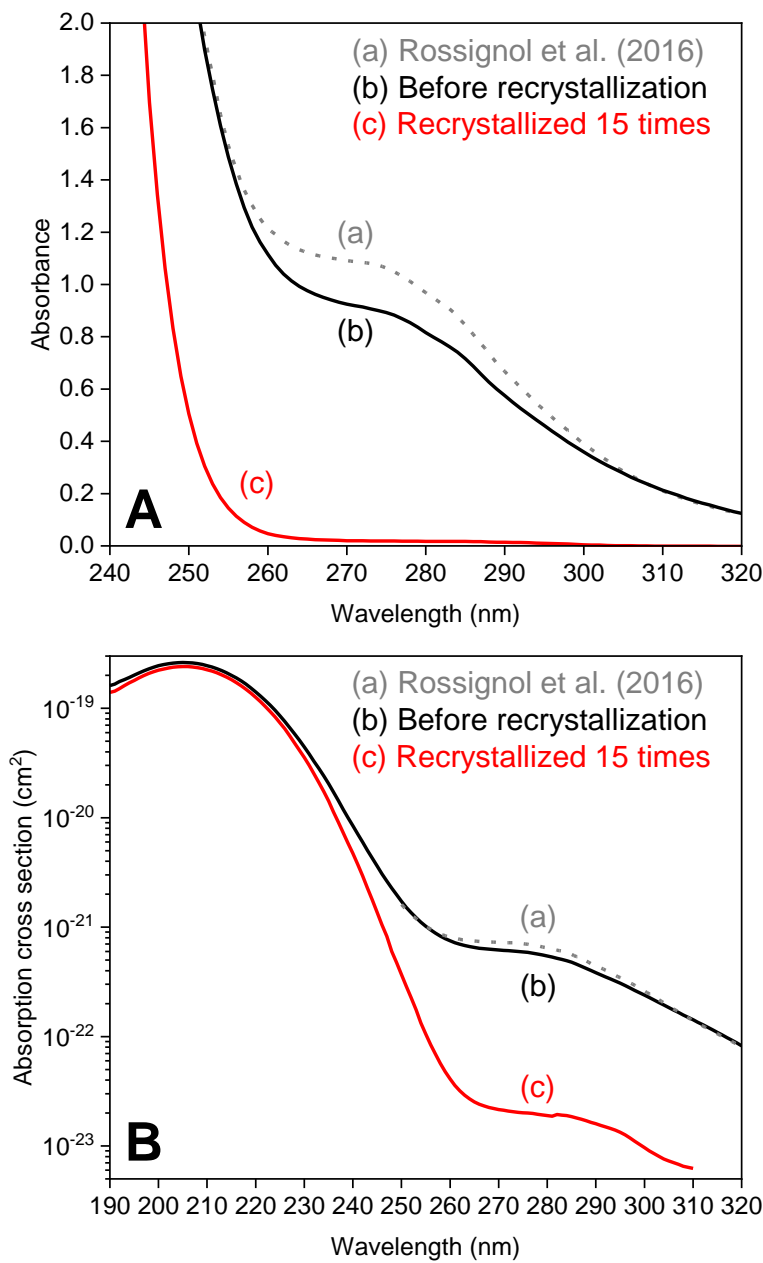

**Fig. 1. UV absorption spectra and absorption cross sections of liquid nonanoic acid at room temperature.** (A) UV absorption spectra of nonanoic acid: (a) reported by Rossignol et al. (*6*) for a commercial sample (purity 97%); measured here for a commercial sample (purity > 98.0%) (b) before and (c) after recrystallization. Optical path lengths were 10 mm. (B) The corresponding calculated absorption cross sections.

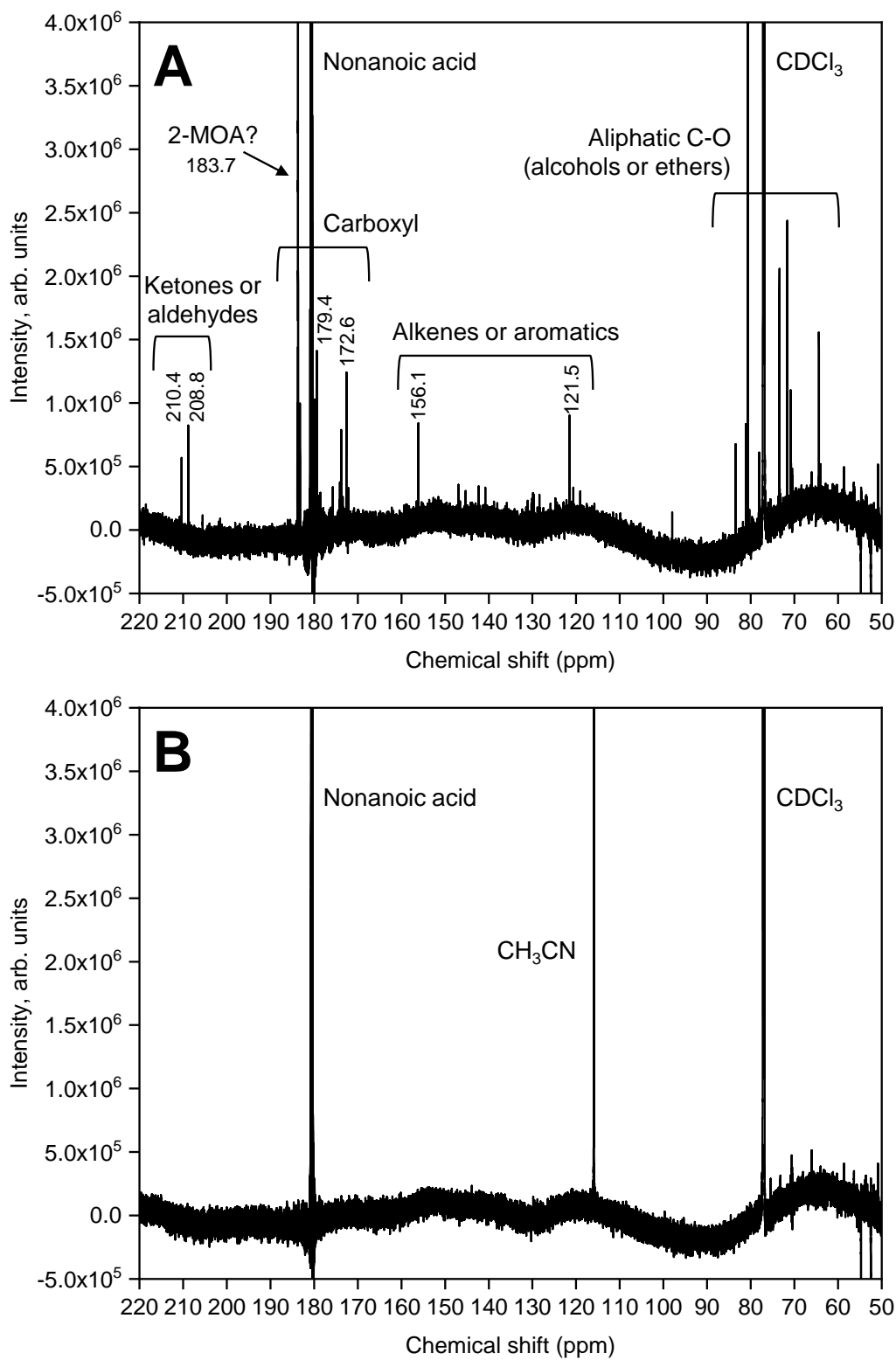

**Fig. 2. $^{13}$C NMR spectra of 50% deuterated chloroform (CDCl$_3$) solutions of nonanoic acid (purity > 98.0%).** (**A**) Before and (**B**) after recrystallization. 2-MOA represents 2-methyloctanoic acid [CH$_3$(CH$_2$)$_4$CH(CH$_3$)COOH] as a candidate for the most abundant impurity.

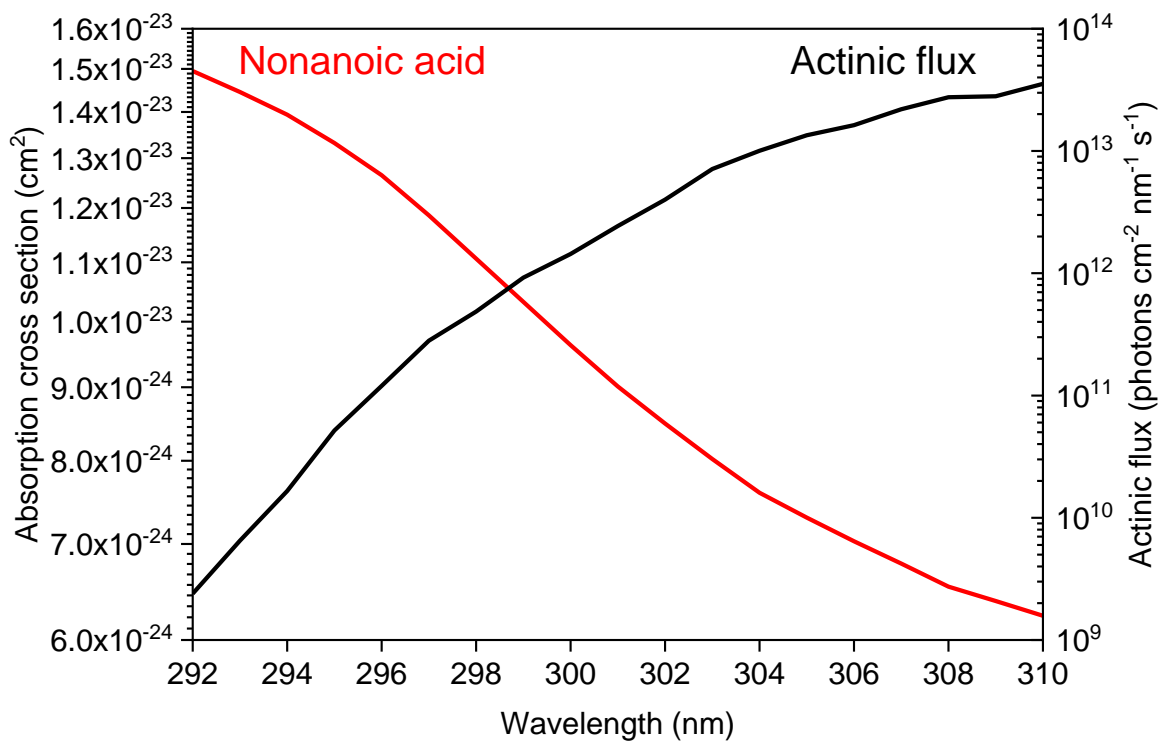

**Fig. 3. Absorption cross section of purified nonanoic acid and actinic flux.** The actinic flux data (measured at 0.1 km altitude and solar zenith angle of 17°) are adopted from reference (*23*).